\documentclass[aps,prb,reprint,superscriptaddress,nofootinbib]{revtex4-2}
\usepackage{graphicx,epsfig}
\usepackage{times}
\usepackage{graphics,dcolumn,bm,float}
\usepackage{amssymb,amsmath,rotate,color,dsfont}
\usepackage[breaklinks,colorlinks=true,urlcolor=blue,citecolor=blue,linkcolor=blue]{hyperref}
\usepackage{breakurl}
\usepackage[T1]{fontenc}

\begin{document}
\unitlength 1 cm
\newcommand{\be}{\begin{equation}}
\newcommand{\ee}{\end{equation}}
\newcommand{\bearr}{\begin{eqnarray}}
\newcommand{\eearr}{\end{eqnarray}}
\newcommand{\nn}{\nonumber}
\newcommand{\dagg}{{\dagger}}
\newcommand{\vpdag}{{\vphantom{\dagger}}}
\newcommand{\vecr}{\vec{r}}
\newcommand{\bs}{\boldsymbol}
\newcommand{\up}{\uparrow}
\newcommand{\dn}{\downarrow}
\newcommand{\fns}{\footnotesize}
\newcommand{\ns}{\normalsize}
\newcommand{\cdag}{c^{\dagger}}
\newcommand{\so}{\lambda_{\rm SO}}
\newcommand{\jh}{J_{\rm H}}
\newcommand{\jk}{J_{\rm K}}

\definecolor{red}{rgb}{1.0,0.0,0.0}
\definecolor{green}{rgb}{0.0,1.0,0.0}
\definecolor{blue}{rgb}{0.0,0.0,1.0}

\title{Antiferromagnetic topological insulators in heavy-fermion systems}

\author{Mohsen Hafez-Torbati}
\email{m.hafeztorbati@gmail.com}
\affiliation{Department of Physics, Shahid Beheshti University, 1983969411, Tehran, Iran}

\begin{abstract}
The cooperation of electronic correlation and spin-orbit coupling can stabilize magnetic topological insulators
which host novel quantum phenomena such as the quantum anomalous Hall state also known as Chern insulator (CI).
Here, we investigate the existence of magnetic topological insulators with antiferromagnetic (AF) order in
heavy-fermion materials.
Our analysis relies on the half-filled Kane-Mele-Kondo (KMK) model with the AF Kondo interaction $\jk$ coupling
the spin of itinerant electrons with a $S=1/2$ localized spin at each lattice site. We consider the N\'eel AF ordering
with the local magnetization not only perpendicular ($z$-AF ordering) but also parallel ($xy$-AF ordering) to
the honeycomb plane. We show that in the absence of an energy offset between the
two sublattices of the honeycomb structure the system is always topologically trivial. There is a transition from the
trivial $xy$-AF insulator ($xy$-AFI) to the trivial Kondo insulator (KI) upon increasing $\jk$.
We unveil that an alternating sublattice potential can
lead to the stabilization of
the $z$-AFCI and the $z$-AF quantum spin Hall insulator ($z$-AFQSHI).
We address the charge excitations in the bulk as well as at the edges of the KMK model.
We provide a systematic comparison between the size of the charge gap in the AFCI in heavy-fermion materials and
the size of the charge gap in the AFCI
in transition-metal compounds.
Our findings can guide the future experimental studies searching for AF topological insulators in novel class of systems
which can survive up to higher temperatures.

\end{abstract}

\maketitle

\section{Introduction}
Since the experimental discovery of the quantum anomalous Hall effect in thin films of the topological
insulator (Bi,Sb)$_2$Te$_3$ doped with the transition-metal element Cr at temperatures below $30$ mK
\cite{Chang2013} there
has been a large interest to stablish the effect in other systems and at higher temperatures
\cite{Tokura2019,Chang2023}.
The higher temperature realization is essential not only for application of the dissipationless charge
transport in spintronic technology but also to extend the possible experimental technique for a deeper
understanding of the effect and its related phenomena. Using high-quality samples \cite{Kou2015} and
a modulation doping technique the Chern insulator phase is realized at temperatures up to 2 K \cite{Mogi2015}.
This is still more than one order of magnitude smaller than the Curie temperature
 $T_{\rm C} \approx 30$ K of the material.

In order to omit the detrimental effect of the disorder and shift the quantization temperature of the Hall
conductance towards the magnetic transition temperature of the material extensive research efforts have been dedicated
in recent years to finding magnetic topological insulators which show intrinsic magnetic ordering \cite{Wang2021,Wang2023}.
This has led to
the theoretical prediction and experimental verification of multiple intrinsic magnetic topological
insulators such as MnBi$_2$Te$_4$ \cite{Otrokov2019a,Otrokov2019b,Li2019a,Gong2019,Li2019b,Hao2019,Chen2019}.
The existence of the Chern insulator \cite{Deng2020} as well as the axion insulator \cite{Liu2020}
phase are confirmed in thin films of MnBi$_2$Te$_4$ depending on the number of septuple layers being odd or even.
Despite the remarkable development in the realization of the CI in an intrinsic magnetic topological
insulator the reported quantization temperature is still limited to below 2 K \cite{Deng2020}.
This can be attributed to the negligible charge gap in the system \cite{Li2019b,Hao2019,Chen2019}.

The third class of systems in which the CI phase is observed is moir\'e materials. This includes twisted bilayer
graphene aligned to hexagonal boron nitride \cite{Serlin2020} and twisted transition metal
dichalcogenide heterobilayers \cite{Li2021}.
In contrast to the magnetically doped topological insulators a charge gap several times the
Curie temperature is reported. However, the small Curie temperature $T_{\rm C}\approx 7$ K still limits
the observation of the quantum anomalous Hall effect to only a few kelvins.

In all the above cases the stabilization of the CI is linked to the ferromagnetic nature of the material.
Antiferromagnets, however, are much more common in nature in contrast to ferromagnets. They create no stray
field, are robust against disturbing magnetic fields, and support a spin dynamics in the THz regime due to
the large exchange interaction.
These features have made AF materials particularly important for the construction of the next-generation
spintronic devices \cite{Jungwirth2016,Baltz2018,Smejkal2018}.
The realization of an AFCI is more cumbersome than its ferromagnetic counterpart.
This is because in a ferromagnetic
state the time-reversal symmetry is truly broken, i.e., it would not be possible to compensate the
effect of the time-reversal transformation by a space group operation. In an AF state, however, such a composite
antiunitary symmetry can exist preventing a finite Chern number \cite{Jiang2018,Ebrahimkhas2022,Guo2023}.
The AFCI is already reported for transition-metal compounds in single-orbital \cite{Jiang2018,Ebrahimkhas2022}
as well as multi-orbital \cite{HafezTorbati2024} models and also in density functional
theory calculations \cite{Guo2023,Wu2023}.

In this paper we investigate the existence of two-dimensional AF topological insulators including
both the AFCI and the AFQSHI in heavy-fermion systems.
It is one of the main aims
of this study to compare the size of the charge gap in the AFCI in heavy-fermion materials and in transition-metal
compounds.
Our analysis relies on the half-filled KMK model using the dynamical mean-field theory (DMFT) technique
\cite{Georges1996}.
The model is relevant for two-dimensional materials with strong spin-orbit coupling and buckled structure
such as silicene, germanene, and stanene \cite{Molle2017} coupled with appropriate magnetic insulators \cite{Tokmachev2022}.
We consider the N\'eel AF ordering with the local magnetization not only perpendicular to the honeycomb
plane ($z$-AF ordering)
but also parallel to it ($xy$-AF ordering) comparing their energies.
In the absence of an energy offset between the two sublattices of the honeycomb structure
we show that the system is a trivial $xy$-AF insulator ($xy$-AFI) which undergoes a transition to the
trivial Kondo insulator (KI) upon increasing the Kondo interaction $\jk$. We uncover that introducing the alternating
sublattice potential $\delta$ can lead to the stabilization of both the $z$-AFCI and the $z$-AFQSHI.
Complementary to the topological invariant calculations we confirm
the existence of the gapless charge excitations localized at the edges for only one spin component in
the AFCI phase and also the closing of the bulk charge gap across the topological phase transitions.
We provide a systematic comparison between the size of the charge gap in the AFCI in heavy-fermion
systems and the size of
the charge gap in the AFCI reported for transition-metal compounds \cite{HafezTorbati2024}.

The paper is organized as follows. In the next section we introduce the KMK model and discuss the technical aspects.
In Sec. \ref{sec:pd} we present the phase diagram of the KMK model in the $\jk$-$\delta$ plane. Sec. \ref{sec:gap} is devoted
to a comprehensive study of the charge excitations in the system.
The paper is concluded in the last section.

\section{Model and Method}
\label{sec:mod}
The KMK model serves as a paradigm for the study of the effect of the spin-orbit coupling
in heavy-fermion systems \cite{Feng2013,Zhong2013,Yoshida2016}.
The Hamiltonian of the model is given by
\begin{align}
H=&+t\sum_{\langle i,j\rangle} \sum_{\alpha} c^{\dagger}_{i\alpha} c^{\vpdag}_{j\alpha}
+{\rm i}\lambda_{\rm SO} \sum_{[ i,j ]}
\sum_{\alpha \beta}
\nu_{ij}^{\vpdag}
c^{\dagger}_{i\alpha} \sigma^{z}_{\alpha\beta} c^{\vpdag}_{j\beta} \nn \\
&+\sum_{i} \sum_{\alpha} \delta_i^{\vpdag}  c^{\dagger}_{i\alpha} c^{\vpdag}_{i\alpha}
+\jk \sum_{i} \vec{s}_i \cdot \vec{S}_i \ ,
\label{eq:kmk}
\end{align}
where $c^{\dagger}_{i\alpha}$ and $c^{\vpdag}_{i\alpha}$ are the usual creation and annihilation
fermionic operators at the honeycomb lattice site labeled by $i$ and with the $z$
component of the electron spin  $\alpha=\up$ or $\downarrow$.
The notation $\langle i,j\rangle$ in the first term and the notation $[ i,j]$ in the second term
limit sites $i$ and $j$ to be nearest-neighbor (NN) and next-nearest-neighbor, respectively.
The first term in Eq. \eqref{eq:kmk} is the NN hopping.
The second term originates from the spin-orbit coupling with the honeycomb
layer chosen in the $x$-$y$ plane.
$\sigma^{z}$ is the Pauli matrix and $\nu_{ij}=2/\sqrt{3}(\hat{d}_1 \times \hat{d}_2)_{z}=\pm 1$ with
$\hat{d}_1$ and $\hat{d}_2$ being unit vectors along the two bonds the electron traverses going from site $j$
to $i$ \cite{Kane2005}. We suppose $\so \geq 0$ without the loss of generality. The third term is an alternating sublattice potential
giving the onsite energy $\delta_i=+\delta$ to the sublattice $A$ and the onsite energy $\delta_i=-\delta$ to the sublattice $B$ of the
honeycomb structure. Such an alternating sublattice potential is extensively used to study transitions between
topological and normal insulators \cite{Kane2005a,Guo2011,Amaricci2015,Hafez-Torbati2020} and also to
realize AF topological states \cite{Jiang2018,Ebrahimkhas2022,Guo2023,Vanhala2016,Ebrahimkhas2021}.
The alternating sublattice potential can be
fine-tuned by applying a perpendicular electric field to a buckled structure \cite{Ezawa2012,Xiao2011}.
The last term in Eq. \eqref{eq:kmk} is the AF Kondo interaction ($\jk>0$) coupling the electron
spin $\vec{s}_i$ to the localized spin $\vec{S}_i$ with the size $S=1/2$ at each lattice site.
The Hamiltonian is relevant for two-dimensional systems with a strong spin-orbit coupling and buckled
structure such as silicene \cite{Molle2017} grown on an appropriate magnetic insulating substrate
\cite{Tokmachev2022}.
The Heisenberg exchange interaction between the localized spins neglected in Eq. \eqref{eq:kmk} is not expected to have a
substantial effect on our results except shifting the transition to the KI state to a larger Kondo
coupling $\jk$.

The first three terms in Eq. \eqref{eq:kmk} define the Kane-Mele model $H_{\rm KM}$ \cite{Kane2005,Kane2005a}.
For $\delta<3\sqrt{3}\lambda_{\rm SO}$ the opposite spins carry the opposite Chern numbers
$\mathcal{C}_{\alpha}={\rm sgn}(\alpha)$ with ${\rm sgn}(\up)=+1$  and ${\rm sgn}(\downarrow)=-1$
and the system describes a QSHI characterized by the
spin Chern number
$\mathcal{C}_s=(\mathcal{C}_{\uparrow}-\mathcal{C}_{\downarrow})/2=1$.
Increasing the value of $\delta$ the charge gap closes at $\delta=3\sqrt{3}\lambda_{\rm SO}$ and
a transition to the trivial band insulator phase with $\mathcal{C}_{\alpha}=0$ takes place.
An infinitesimal Kondo interaction couples the electron spins to the localized spins
and leads to the magnetic ordering due to the van Vleck mechanism \cite{Vleck1932,Bloembergen1955}.
In the opposite limit that
$\jk$ is the dominant term the electron spin and the localized spin at each lattice
site form a singlet state and the system becomes a paramagnetic KI.

The third interesting limit is the atomic limit for which $t=\so=0$. For $\delta>3\jk/4$ the sublattice with
the lower onsite energy is fully occupied and the sublattice with the higher onsite energy is empty.
For $\delta<3\jk/4$ each lattice site is occupied by one electron forming a singlet with the localized spin.
These two states with completely different charge
distributions become degenerate at $\delta=3\jk/4$. Exotic quantum phases are expected to emerge around
$\delta = 3\jk/4$ upon introducing the hopping and the spin-orbit coupling.

\begin{figure}[t]
   \begin{center}
   \includegraphics[width=0.44\textwidth,angle=0]{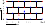}
   \caption{The treatment of the honeycomb lattice as a brick wall.
   The number of lattice sites in the $x$ direction is $L_x$ and in the $y$ direction is $L_y$.
   Each lattice site is identified by two integers $(x,y)$.
   The higher energy sublattice $A$ and the lower energy sublattice $B$ are specified.}
   \label{fig:bw}
   \end{center}
\end{figure}

To address the KMK model from weak to strong interaction limits we adopt the DMFT
which is an established method for strongly correlated systems \cite{Georges1996,Hofstetter2018}
and is widely used to investigate the interplay of strong electronic correlation
and spin-orbit coupling
\cite{Rachel2018,Yu2011,Yoshida2016,Amaricci2015,Cocks2012,Budich2013,Ebrahimkhas2022}.
The method fully takes into account the local quantum fluctuations but neglect
non-local quantum fluctuations by approximating the self-energy to be
spatially local, $\Sigma_{ij;\alpha\beta}({\rm i}\omega_n)=\Sigma_{i;\alpha\beta}({\rm i}\omega_n)\delta_{ij}$.
The spin off-diagonal elements of the self-energy corresponding to $\alpha \neq \beta$ are necessary
to access solutions with a local magnetization out-of the $z$ direction. Note that addressing these solutions
are necessary for a proper analysis of the model as a finite spin-orbit coupling in Eq. \eqref{eq:kmk}
reduces the SU(2) symmetry to the U(1) symmetry.
We employ the real-space realization of the DMFT \cite{Potthoff1999,Song2008,Snoek2008}
as it allows to capture not only the bulk but also the edge
properties on equal footing. We specifically use the implementation introduced in Ref. \onlinecite{Hafez-Torbati2018}
which is suitable for capturing different types of magnetic orderings. It is already applied
to Kondo lattice models with \cite{HafezTorbati2024} and without spin-orbit coupling
\cite{Hafez-Torbati2021,Hafez-Torbati2022}.
The lattice model is mapped to a set of effective Kondo impurity problems depending on the
symmetry of the lattice. For bulk properties using periodic boundary conditions in both directions
two Kondo impurity models are set up, one for sublattice $A$ and one for sublattice $B$. In the absence
of the sublattice potential $\delta$ this can be reduced to one even in the N\'eel AF phase
because the phase is still symmetric with respect to a combined swap of the sublattice and the spin orientation.
To address edge properties we consider open boundary
conditions in $x$ and periodic boundary conditions in $y$ direction (cylindrical geometry)
treating the honeycomb structure as a brick wall shown in Fig. \ref{fig:bw} with each lattice site labeled by
two integers $x$ and $y$. We have mainly used $L\times L$ lattices with $L=40$ but we have produced
data also for $L=60$ for selective points close to phase transitions making sure that the results are
independent than the system size.

We opt for exact diagonalization (ED) \cite{Georges1996,Caffarel1994}
to solve the Kondo impurity problem because it permits for  a quantum
mechanical treatment of the localized spin in contrast to the quantum Monte Carlo (QMC) solver which
suffers from the fermionic sign problem and is restricted only to classical spins \cite{Yunoki1998,Peters2006,Motome2003}.
In addition, the ED solver provides a direct access to the real-frequency
spectral function while in a QMC treatment an analytical continuation is required. Due to the finite number
of bath sites in the ED solver the spectral function consists of separate sharp peaks which approximate a
continuous function. However, the charge gap extracted from the ED spectral function is found accurate
and is used to benchmark the results obtained from the QMC solver \cite{Wang2009}.
In view of a future finite temperature
analysis we perform a full diagonalization for comparability. We provide
data for different number of bath sites corroborating the accuracy of the results.

To distinguish distinct topological phases one needs to compute topological invariants, which is challenging
for an interacting model. The topological invariant of an interacting system can be computed using twisted
boundary conditions \cite{Niu1985}.
However, it requires access to the eigenstates of the Hamiltonian which is out of the DMFT scope.
In the absence of a non-trivial ground state degeneracy, such as the one occurring in fractional
quantum Hall effect, the topological invariant of an interacting system can be computed using
the Ishikawa-Matsuyama formula which requires momentum and frequency integration over the Green's
function and its derivatives \cite{Ishikawa1987,Wang2010,Gurarie2011}.
Using an adiabatic deformation of the Green's function without the charge gap closing it is shown that
only the zero-frequency Green's function is needed to determine the topological invariant of an
interacting model \cite{Wang2012}. The approach is called the topological Hamiltonian method.
It relates the topological
invariant of an interacting system to the topological invariant of an effective non-interacting model
known as the topological Hamiltonian which in the second quantization form is given by
\be
H_{\rm top}=H_0+\sum_{ij}\sum_{\alpha\beta} c^{\dagger}_{i\alpha} \Sigma_{ij;\alpha\beta} (0)  c^{\vpdag}_{j\beta} \quad,
\label{eq:th}
\ee
where $H_0$ stands for the non-interacting part of the original Hamiltonian and $\Sigma_{ij;\alpha\beta} (0)$
is the self-energy at zero frequency.

The topological Hamiltonian approach should be used with caution.
The derivation of the topological Hamiltonian  \eqref{eq:th} relies on the adiabatic deformation of the
imaginary frequency Green's function such that it never becomes singular, i.e., the charge gap never closes.
However, the Green's function and its inverse appear symmetrically in the Ishikawa-Matsuyama formula and
the topological invariant can change not only if the poles but also if the zeroes of the Green's function
occur \cite{Blason2023}.
The latter corresponds to the divergence of the self-energy and can be captured in our DMFT analysis.

In the DMFT approximation the self-energy is local. In the $z$-AF phase the self-energy in addition
has no off-diagonal spin element,
$\Sigma_{i;\alpha\beta}({\rm i}\omega_n)=\delta_{\alpha\beta}\Sigma_{i;\alpha}({\rm i}\omega_n)$,
and is real at zero frequency. Using the symmetry $\Sigma_{A;\alpha}(0)=-\Sigma_{B;\alpha}(0)$ between the zero-frequency
self-energies on sublattices $A$ and $B$ the topological Hamiltonian simplifies to the Kane-Mele model with the
effective spin-dependent alternating sublattice potential
\be
\tilde{\delta}_{\alpha}=\delta+\Sigma_{A;\alpha}(0) \quad,
\label{eq:edelta}
\ee
where $A$ is the sublattice with the higher onsite energy.
Using this relation one can easily find out
the Chern number for each spin component and determine the topological nature of a $z$-AF state.
The spin component $\alpha$ has the Chern number $\mathcal{C}_\alpha={\rm sgn}(\alpha)$ for
$|\tilde{\delta}_{\alpha}|<3\sqrt{3}\so$ and the Chern number $\mathcal{C}_\alpha=0$ for
$|\tilde{\delta}_{\alpha}|>3\sqrt{3}\so$.
The $z$-AFCI is characterized by the total Chern number $\mathcal{C}=\mathcal{C_\uparrow}+\mathcal{C_\downarrow}\neq 0$.
In the $z$-AFQSHI the total Chern number is zero but there is a finite spin Chern number
$\mathcal{C}_s=(\mathcal{C_\uparrow}-\mathcal{C_\downarrow})/{2}$ mod $2$.
The trivial $z$-AFI corresponds to $\mathcal{C}=\mathcal{C}_s=0$.

In the absence of the alternating sublattice potential, $\delta=0$, there is the additional symmetry
$\Sigma_{A;\alpha}(0)=\Sigma_{B;\bar{\alpha}}(0)$ between the spin-up and the spin-down sublattices
of the N\'eel AF state, where $\bar{\alpha}$ denotes the opposite direction of $\alpha$. This leads to
$\tilde{\delta}_{\up}=-\tilde{\delta}_{\downarrow}$ which excludes the possibility of the emergence of the AFCI.
This originates from the fact that for $\delta=0$ the electronic state is invariant under the combination
of the time-reversal and the inversion transformations \cite{Jiang2018,Ebrahimkhas2022,Guo2023}.

In the $xy$-AF state the spin diagonal elements obey the spin symmetry
$\Sigma_{i;\uparrow\uparrow}(0)=\Sigma_{i;\downarrow\downarrow}(0)=:\Sigma_{i}(0)\in \mathbb{R}$ and the sublattice symmetry
$\Sigma_{A}(0)=-\Sigma_{B}(0)$. For the spin off-diagonal elements one has
$\Sigma_{i;\uparrow\downarrow}(0)=\Sigma^*_{i;\downarrow\uparrow}(0)$ and
$\Sigma_{A;\uparrow\downarrow}(0)=-\Sigma_{B;\uparrow\downarrow}(0)$.
This simplifies the topological Hamiltonian \eqref{eq:th} to
the Kane-Mele model with the effective spin-independent sublattice potential
$\tilde{\delta}=\delta+\Sigma_{A}(0)$ subjected to the effective staggered in-plane magnetic field
\be
\label{eq:eh}
\vec{h}_{A}=-\vec{h}_{B}=2(-{\rm Re}[\Sigma_{A;\uparrow\downarrow}(0)],
+{\rm Im}[\Sigma_{A;\uparrow\downarrow}(0)],~0)^{\top} \ ,
\ee
with $\vec{h}_{A}$ and $\vec{h}_{B}$ acting respectively on the sublattice $A$ and on the sublattice $B$.
An infinitesimal staggered in-plane magnetic
field \eqref{eq:eh} makes the edge states gapped and the system becomes a trivial insulator \cite{Zhu2023}.
Hence, the $xy$-AF phase is always topologically trivial.

In the limit of large $\jk$ the system is in the paramagnetic KI state. The self-energy has no spin off-diagonal
element and the diagonal elements are spin-independent. At zero frequency the self-energy is real and obeys the sublattice symmetry
$\Sigma_{A}(0)=-\Sigma_{B}(0)$. The topological Hamiltonian
is given by the Kane-Mele model with the effective alternating sublattice potential $\tilde{\delta}=\delta+\Sigma_{A}(0)$.
This is just a special case of the above topological Hamiltonians derived in
the magnetically ordered phases. We find that
the zero-frequency self-energy in the KI phase is always very large such that $|\tilde{\delta}|>3\sqrt{3}\so$ and
hence the KI always falls in the topologically trivial region.
It should be mentioned that in the KI phase
the self-energy at zero frequency diverges (zeroes in the Green's function occur)
for $\delta\to 0$ which is the responsible mechanism for the
gap opening. It is demonstrated by direct evaluation of the Ishikawa-Matsuyama
formula that the KI phase at $\delta= 0$ is topologically trivial \cite{Yoshida2016}.

The combination of the DMFT and the topological Hamiltonian method is widely used to map out the phase diagram
of different interacting topological models \cite{Vanhala2016,Budich2013,Ebrahimkhas2021,Irsigler2019,
Amaricci2015,Hafez-Torbati2020,Gu2019}.
The result obtained for the phase diagram of the Haldane-Hubbard model is
in qualitative agreement with the results obtained by other methods such as the ED \cite{Vanhala2016}
and the density-matrix renormalization group \cite{He2024}.
A systematic treatment of the non-local quantum fluctuations beyond the DMFT also indicates
only miner changes to the phase boundaries \cite{Mertz2019}.
Accordingly, we believe our results to be qualitatively reliable and including the non-local quantum
fluctuations to have only miner effects on the phase boundaries.

We emphasize that our characterization of different topological phases does not rely solely on
the topological Hamiltonian method. We address the charge excitations in the bulk and at the edges directly for
the interacting model \eqref{eq:kmk}. We show that there is a finite charge gap in the bulk and at
the edges for topologically trivial phases while in the AFCI phase there are gapless charge excitations
localized at the edges for only one spin component. Moreover, we show the tendency for the charge gap closing
in the KMK model at the location of the topological transition points predicted by the topological Hamiltonian.

\section{Phase diagram}
\label{sec:pd}

\begin{figure}[t]
   \begin{center}
   \includegraphics[width=0.46\textwidth,angle=0]{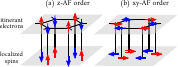}
   \caption{Schematic representation of the $z$-AF order (a) and the $xy$-AF order (b) in the Kane-Mele-Kondo model
   with an AF Kondo interaction between the spin of itinerant electrons and the localized spins.
   The red filled circles stand for the sites with the higher onsite energy and the
   blue filled circles stand for the sites with the lower onsite energy.
   The $xy$-AF state
   is continuously degenerate due to the spontaneous breaking of the U(1) symmetry.
   The $z$-AF state is two-fold degenerate due to the spontaneous breaking of the time-reversal
   symmetry.}
   \label{fig:af}
   \end{center}
\end{figure}

In this section we map out the $T=0$ phase diagram of the KMK model \eqref{eq:kmk} in the $\jk$-$\delta$ plane for the spin-orbit
coupling $\so=0.2t$. To distinguish the different phases we compute three different quantities. The magnitude of the
local magnetization
of itinerant electrons $m:=|\langle \vec{s}_i \rangle|$ and the localized spins $M:=|\langle \vec{S}_i \rangle|$ pinpoint the
transition from the magnetically ordered phase to the paramagnetic KI state. One notes that the lattice always has the
N\'eel AF order and at each lattice site the spin of itinerant electrons $\langle \vec{s}_i \rangle$ and the localized spin
$\langle \vec{S}_i \rangle$ are antiparallel. We consider N\'eel AF solutions with the local magnetization not
only perpendicular to the honeycomb plane, $z$-AF solution, but also parallel to it, $xy$-AF solution.
A schematic representation of the $z$-AF order and the $xy$-AF order is given in Fig. \ref{fig:af}.
We present data
for their energy difference per lattice site, $\varepsilon_z-\varepsilon_{xy}$, to determine which solution is the ground
state. Finally, the effective alternating sublattice potential \eqref{eq:edelta} reveals the topological nature of the
$z$-AF state. The $xy$-AF phase and the KI phase are always topologically trivial as we discussed in the previous section.
We first focus on the case $\delta=0$ and then consider the effect of a finite $\delta$.

\begin{figure}[t]
   \begin{center}
   \includegraphics[width=0.57\textwidth,angle=-90]{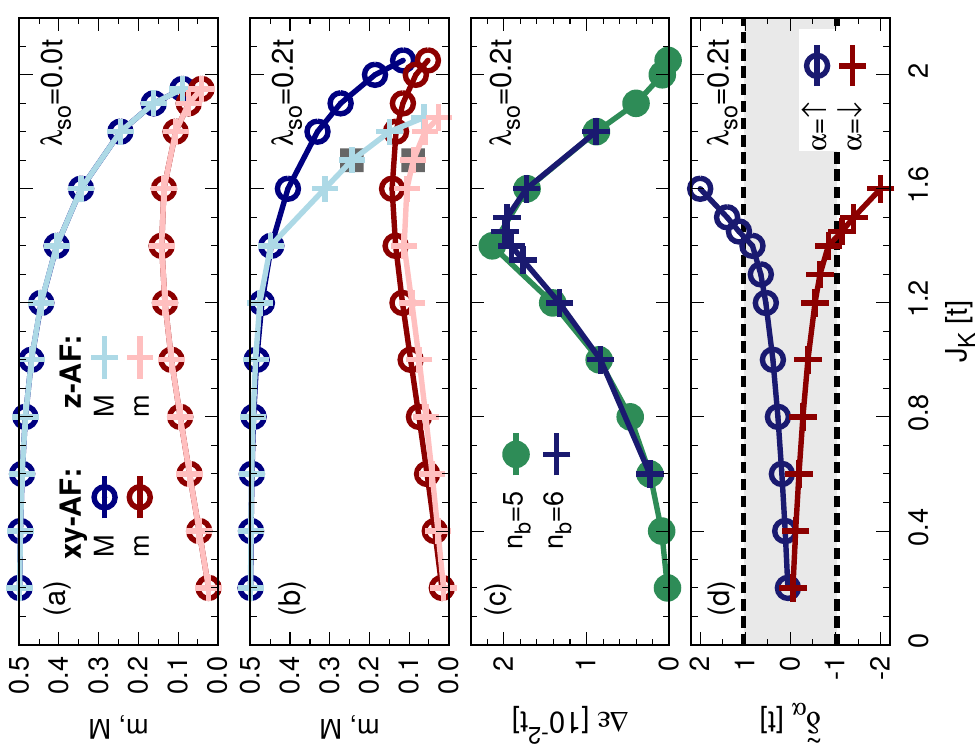}
   \caption{The results for Kane-Mele-Kondo model \eqref{eq:kmk} at the alternating sublattice potential $\delta=0$.
   (a),(b) The magnitude of the local magnetization of the itinerant electrons $m$ and the localized spins $M$
   in the $z$-AF solution and in the $xy$-AF solution vs the Kondo interaction $\jk$ for
   the spin-orbit coupling $\so=0$
   in (a) and $\so=0.2t$ in (b).
   (c) The energy difference per lattice site between the $z$-AF solution and the $xy$-AF solution,
   $\Delta\varepsilon=\varepsilon_z-\varepsilon_{xy}$, vs $\jk$ for $\so=0.2t$.
   For $\jk > 1.85t$ the $z$-AF solution in panel (b) undergoes a transition to the KI and accordingly panel (c) for $\jk > 1.85t$
   exhibits the energy difference between the KI state and the $xy$-AF state.
   (d) The effective alternating sublattice potential \eqref{eq:edelta} vs $\jk$ for $\so=0.2t$. The dashed lines
   separate the topological region (solid area) from the trivial region. The results in panels (a), (b), and (d)
   are for the number of bath sites $n_b=5$ in the ED impurity solver except for the filled gray squares in panel
   (b) at $\jk=1.7t$ which are for $n_b=7$ in the $z$-AF solution.}
   \label{fig:delta0}
   \end{center}
\end{figure}

Fig. \ref{fig:delta0} represents the results for the KMK model at $\delta=0$. In panel (a) we have plotted the local
magnetization of the itinerant electrons $m$ and the localized spins $M$ vs $\jk$ in the $z$-AF solution and in the
$xy$-AF solution in the absence of the spin-orbit coupling, $\so=0$. As one can see the results for the two solutions
perfectly match. This is the case also for the energy that we find for the two solutions and
is expected because for $\so=0$ the Hamiltonian \eqref{eq:kmk} reduces to the Kondo lattice model which obeys the SU(2)
symmetry. This serves as a test for the solutions.

Upon introducing the spin-orbit coupling $\so=0.2t$ in Fig. \ref{fig:delta0}(b) the results for the
$z$-AF solution and the $xy$-AF solution no longer coincide. The $xy$-AF solution persists to larger
values of the Kondo interaction while the $z$-AF solution closes at a smaller value of $\jk$ in contrast
to the results in Fig. \ref{fig:delta0}(a).
The data are for the number of bath sites $n_b=5$ except for the filled gray squares at $\jk=1.7t$
which are for $n_b=7$ for the $z$-AF solution corroborating the accuracy of the results.
The energy difference per lattice site between the $z$-AF solution and the $xy$-AF solution,
$\varepsilon_z-\varepsilon_{xy}$, depicted
in Fig. \ref{fig:delta0}(c) reveals that the $xy$-AF solution always has the lower energy.
The results given for the different number of bath sites $n_b=5$ and $6$ indicate the accuracy of
the data. There is only a small difference between the data for $n_b=5$ and $6$ near the peak of the graph at
$\jk \approx 1.4t$ otherwise they perfectly match. One notes that for $\jk > 1.85t$ the $z$-AF solution
in Fig. \ref{fig:delta0}(b) undergoes a transition to the KI and accordingly Fig. \ref{fig:delta0}(c) for $\jk > 1.85t$
exhibits the energy difference between the KI state and the $xy$-AF state. The results in Figs. \ref{fig:delta0}(b)
and \ref{fig:delta0}(c)
unveil that the KMK model at $\delta=0$ displays only a single phase transition
from the topologically trivial $xy$-AFI to the topologically trivial KI upon increasing $\jk$.
No non-trivial topological phase stabilizes.
We would like to mention that
we have produced not one, but multiple $xy$-AF solutions for some selective points checking that
they all display the same magnitude of the local magnetization and the same energy as is expected
from the spontaneous breaking of the U(1) symmetry.

The half-filled KMK model \eqref{eq:kmk} at $\delta=0$ is investigated in a number of previous studies focusing on
the paramagnetic solutions only \cite{Feng2013} but including also the AF solutions \cite{Zhong2013,Yoshida2016}.
The AF solutions, however, are still restricted
to only the $z$-AF ordering. The possibility for the $xy$-AF ordering is not taken into account.
Two quantum phase transitions are identified upon increasing the Kondo interaction $\jk$. The first is
from the AFQSHI to the trivial AFI and the second is from the trivial AFI to the KI.
Our results for the local magnetization in Fig. \ref{fig:delta0}(b) and the effective alternating sublattice
potential \eqref{eq:edelta} plotted in Fig. \ref{fig:delta0}(d) indeed confirms this picture for the $z$-AF solution.
The dashed lines at $|\tilde{\delta}_{\alpha}|=3\sqrt{3}\so$ in Fig. \ref{fig:delta0}(d) separate the topological region
(solid area) from the trivial region. The transition from the $z$-AFQSHI to the trivial $z$-AFI
occurs at $J_{{\rm K}} \approx 1.4t$ where $\tilde{\delta}_{\up}=-\tilde{\delta}_{\dn}$ crosses the dashed line.
The transition point found at the smaller value $J_{{\rm K}} \approx 0.7t$ in Ref. \onlinecite{Yoshida2016}
is due to the smaller spin-orbit coupling $\so=0.1t$ used, which is half the value $\so=0.2t$ used in our calculations.
It should also be noted that in Ref. \onlinecite{Yoshida2016} the numerical renormalization group is employed
as the impurity solver while our analysis relies on the ED solver.
Hence, while our results for the $z$-AF solution agree with the
previous studies we find that the stable phase is the topologically trivial $xy$-AFI
which undergoes a direct transition to the KI upon increasing $\jk$.

In addition to hosting the AFQSHI the $z$-AF order is intriguing because it exhibits
the restoration of topological properties at finite temperatures \cite{Yoshida2016}. Even
the topologically trivial $z$-AFI manifests non-trivial topological properties as the
temperature is raised.
Although our findings exclude the possibility of the $z$-AF ordering at $\delta=0$
one might ask if such a phase can still be realized under some circumstances. In the following
we uncover that introducing the alternating sublattice potential $\delta$ can stabilize the
$z$-AF order, and show that not only the AFQSHI but also the AFCI phase emerge.

\begin{figure}[t]
   \begin{center}
   \includegraphics[width=0.44\textwidth,angle=-90]{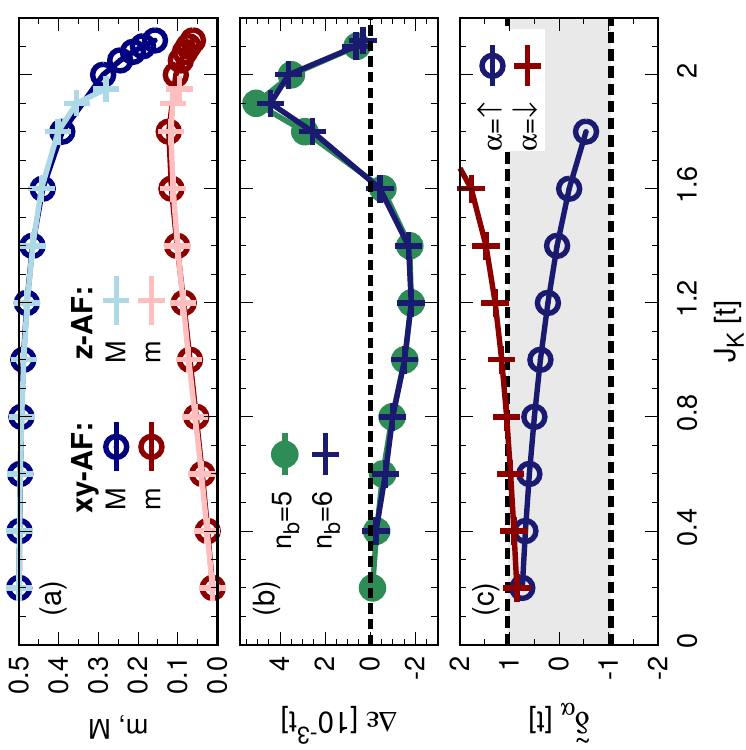}
   \caption{The results for the Kane-Mele-Kondo model \eqref{eq:kmk} with the alternating sublattice potential $\delta=0.8t$
   and the spin-orbit coupling $\so=0.2t$. (a) The magnitude of the local magnetization of the itinerant electrons $m$ and the
   localized spins $M$ in the $z$-AF solution and in the $xy$-AF solution vs the Kondo interaction $\jk$.
   (b) The energy difference per lattice site
   between the $z$-AF solution and the $xy$-AF solution, $\Delta\varepsilon=\varepsilon_z-\varepsilon_{xy}$,
   vs $\jk$. For $\jk > 1.95t$ the $z$-AF solution does not exist and panel (b) compares the energy difference between
   the Kondo insulator (KI) solution and the $xy$-AF solution instead.
   (c) The effective alternating sublattice potential \eqref{eq:edelta} vs $\jk$. The dashed lines separate the
   topological region (solid area) from the trivial region. The results in panels (a) and (c) are for the number of bath sites
   $n_b=5$ in the ED impurity solver.}
   \label{fig:delta0.8}
   \end{center}
\end{figure}

Fig. \ref{fig:delta0.8} displays the results for the KMK model \eqref{eq:kmk} at $\delta=0.8t$ and the
spin-orbit coupling $\so=0.2t$. For these parameters $\delta<3\sqrt{3}\so$ and the Kane-Mele
model lies in the QSHI region. The results for the local magnetization of the itinerant electrons $m$
and the localized spins $M$ in the $z$-AF solution and in the $xy$-AF solution plotted in Fig. \ref{fig:delta0.8}(a)
vs the Kondo interaction indicate that the $xy$-AF phase still persists up to larger values of $\jk$ in contrast
to the $z$-AF phase. We find no $z$-AF solution for $\jk>1.95t$.
The energy difference between the $z$-AF solution and the $xy$-AF solution per lattice site
$\Delta \varepsilon=\varepsilon_z -\varepsilon_{xy}$
in Fig. \ref{fig:delta0.8}(b) reveals that for small $\jk$ it is the $z$-AF solution which has the lower
energy. As $\jk$ is increased the energy difference $\Delta \varepsilon$ crosses the zero-energy line (denoted
as a black dashed line)
at $\jk \approx 1.6t$ and a spin-flop transition to the $xy$-AF phase takes place.
One notes that  for $\jk>1.95t$ the $z$-AF solution no longer exists and Fig. \ref{fig:delta0.8}(b)
compares instead the energy of the KI solution and the $xy$-AF solution,
$\Delta \varepsilon=\varepsilon_{\rm KI}^{\vpdag} -\varepsilon_{xy}$.
The data given for the two different numbers of bath sites $n_b=5$ and $6$ in Fig. \ref{fig:delta0.8}(b)
confirm the accuracy of the results.

To determine the topological nature of the $z$-AF phase the effective alternating sublattice potential
\eqref{eq:edelta} is plotted vs $\jk$ in Fig. \ref{fig:delta0.8}(c). The black dashed lines separate the
topological region $|\tilde{\delta}_\alpha|<3\sqrt{3}\so$ from the trivial
region $|\tilde{\delta}_\alpha|>3\sqrt{3}\so$. For small values of $\jk$ both $\up$ and
$\dn$ spins fall in the topological region, specified by the solid area, and the system is a $z$-AFQSHI
characterized by the spin Chern number $\mathcal{C}_s=(\mathcal{C}_\up-\mathcal{C}_\dn)/2=1$.
Upon increasing $\jk$ one spin (spin $\dn$ in the figure) leaves the topological region at $\jk \approx 0.8t$
while the other remains topological.
This leads to a non-zero total Chern number $\mathcal{C}=\mathcal{C}_\up+\mathcal{C}_\dn$ and the stabilization
of a $z$-AFCI. One notes that a $z$-AF state is two-fold degenerate due to the spontaneous breaking of the time-reversal
symmetry. The results given in Fig. \ref{fig:delta0.8}(c) correspond to the $z$-AF solution with the local
magnetization of the itinerant electrons $\langle s_i^z \rangle$ positive on the higher-energy sublattice $A$
and negative on the lower-energy sublattice $B$.

\begin{figure}[t]
   \begin{center}
   \includegraphics[width=0.27\textwidth,angle=-90]{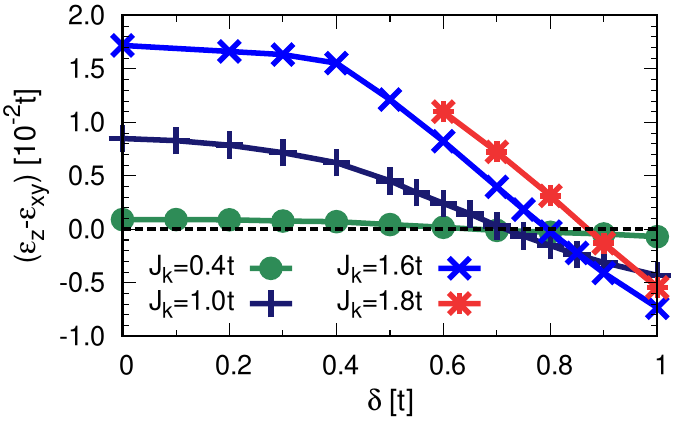}
   \caption{The energy difference per lattice site between the $z$-AF state and the $xy$-AF state
   vs the alternating sublattice potential $\delta$ for various values of the Kondo interaction $\jk$.
   The spin-orbit coupling is fixed to $\so=0.2t$ and the number of bath sites in the ED impurity solver
   is given by $n_b=5$.}
   \label{fig:energy}
   \end{center}
\end{figure}

Overall, the results in Fig. \ref{fig:delta0.8} support three consecutive phase transitions upon increasing $\jk$.
The first is from the $z$-AFQSHI to the $z$-AFCI at $\jk \approx 0.8t$ revealed in Fig. \ref{fig:delta0.8}(c).
The $z$-AFCI persists up to $\jk \approx 1.6t$ where the spin-flop transition to the $xy$-AFI
in Fig. \ref{fig:delta0.8}(b) occurs. The $xy$-AFI undergoes the third phase transition to the KI at
$\jk \approx 2.15t$ as can be seen from Fig. \ref{fig:delta0.8}(a).

The energy difference between the $z$-AF phase and the $xy$-AF phase per lattice site for $\delta=0.8t$
in Fig. \ref{fig:delta0.8}(b) indicates a rather complex behavior in contrast to the results for $\delta=0$
in Fig. \ref{fig:delta0}(c). For $\delta=0$ the $xy$-AF phase always has the lower energy while for
the finite value of the alternating sublattice potential $\delta=0.8t$ it is the $z$-AF phase which
becomes the ground state if $\jk$ is not too large. To see more apparently how the alternating sublattice potential
$\delta$ renders the stabilization of the $z$-AF phase we have plotted in Fig. \ref{fig:energy} the energy difference
$\varepsilon_z-\varepsilon_{xy}$ vs $\delta$ for various values of the Kondo interaction $\jk$. The spin-orbit
coupling $\so=0.2t$ and the number of bath sites $n_b=5$. One can see that for small values of $\delta$
the $xy$-AF phase is always the stable solution. However, upon increasing $\delta$ a spin-flop transition
to the $z$-AF phase takes place. For the Kondo interaction $\jk \lesssim 1.0t$ the spin-flop transition
almost always occurs at the constant value $\delta \approx 0.7t$. As the Kondo interaction is increased
beyond $\jk \approx 1.0t$ the position of the spin-flop transition starts shifting to larger values of $\delta$.
The spin-flop transition found at $\jk \approx 1.6t$ in Fig. \ref{fig:delta0.8}(b) for $\delta=0.8t$ upon
varying $\jk$ is nicely in agreement with the spin-flop transition obtained at $\delta \approx 0.8t$
in Fig. \ref{fig:energy} for $\jk=1.6t$ upon varying $\delta$.

\begin{figure}[t]
   \begin{center}
   \includegraphics[width=0.46\textwidth,angle=-90]{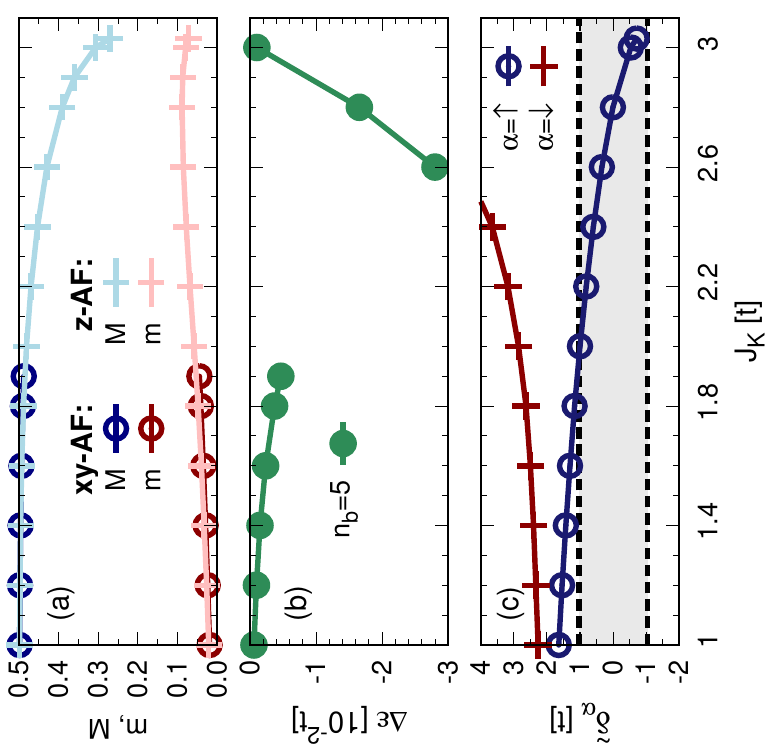}
   \caption{The results for the Kane-Mele-Kondo model \eqref{eq:kmk} with the alternating sublattice potential $\delta=2t$
   and the spin-orbit coupling $\so=0.2t$. (a) The magnitude of the local magnetization of the itinerant electrons $m$ and the
   localized spins $M$ in the $z$-AF solution and in the $xy$-AF solution vs the Kondo interaction $\jk$.
   (b) The energy difference between the $z$-AF solution and the $xy$-AF solution, $\Delta\varepsilon=\varepsilon_z-\varepsilon_{xy}$,
   for $\jk \leq 1.9t$ and between the $z$-AF solution and the Kondo insulator solution for $\jk \geq 2.6t$. For $1.9t <\jk < 2.6t$
   the $z$-AF solution is the only solution which is why there is no data in this parameter range in panel (b).
   (c) The effective alternating sublattice potential \eqref{eq:edelta} vs $\jk$. The dashed lines separate the
   topological region (solid area) from the trivial region. The results are for the number of bath sites
   $n_b=5$ in the ED impurity solver.}
   \label{fig:delta2}
   \end{center}
\end{figure}

We proceed to the rather large value of the alternating sublattice potential $\delta=2t$. For this value
of $\delta$ the Kane-Mele model falls into the trivial insulator phase. Whence, a topologically trivial
AF insulator is expected for weak Kondo interactions. In Fig. \ref{fig:delta2} we have plotted the same
quantities as in Fig. \ref{fig:delta0.8} but for the alternating sublattice potential $\delta=2t$.
In contrast to Figs. \ref{fig:delta0}(b) and \ref{fig:delta0.8}(a) it is the $z$-AF solution in
Fig. \ref{fig:delta2}(a) which persists to larger values of the Kondo interaction $\jk$.
We find a discontinuous transition from the $z$-AF phase to the KI at $\jk \approx 3.05t$.
Fig. \ref{fig:delta2}(b) compares the energy of the $z$-AF solution and the
$xy$-AF solution, $\Delta\varepsilon=\varepsilon_z-\varepsilon_{xy}$, for $\jk \leq 1.9t$ and
the energy of the $z$-AF solution and the paramagnetic KI solution,
$\Delta\varepsilon=\varepsilon_z-\varepsilon_{\rm KI}$, for $\jk \geq 2.6t$.
The $xy$-AF solution exists up to $\jk \approx 1.9t$ and the KI solution exists down to $\jk \approx 2.6t$.
For $1.9t<\jk<2.6t$ the $z$-AF solution is the only solution, which is why there is no data in this
parameter range in Fig. \ref{fig:delta2}(b). The results in Fig. \ref{fig:delta2}(b) reveal that for $\delta=2t$
the $z$-AF solution is always the stable phase in the coexistence regions.

The effective alternating sublattice potential \eqref{eq:edelta} plotted in Fig. \ref{fig:delta2}(c)
unveils a transition from the topologically trivial $z$-AFI to the $z$-AFCI at $\jk \approx 2t$,
which corresponds to the spin $\up$ entering the topological region (solid area).
Similar to the results in Fig. \ref{fig:delta0.8}(c) the results in Fig. \ref{fig:delta2}(c)
are for the $z$-AF solution with the local magnetization of the itinerant electrons $\langle s_i^z \rangle$
positive on the higher-energy sublattice $A$ and negative on the lower-energy sublattice $B$. Applying
the time-reversal transformation one can access the other solution, for which
it is the spin $\dn$ that enters the topological region. The system stays in the $z$-AFCI up to
$\jk \approx 3.05t$ where the transition to the KI takes place, see Fig. \ref{fig:delta2}(a).
We have produced data also for larger values of $\delta$ and found
similar phase transitions except that occurring at larger values of $\jk$.

\begin{figure}[t]
   \begin{center}
   \includegraphics[width=0.32\textwidth,angle=-90]{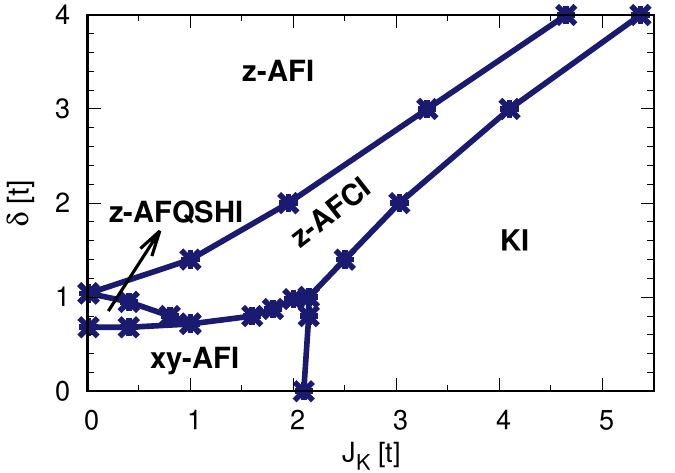}
   \caption{The phase diagram of the KMK model \eqref{eq:kmk} for the spin-orbit coupling $\so=0.2t$.
   The topologically trivial phases are the $z$-AF insulator ($z$-AFI), the $xy$-AF insulator ($xy$-AFI),
   and the Kondo insulator (KI). The non-trivial topological phases are the $z$-AF Chern insulator ($z$-AFCI)
   and the $z$-AF quantum spin Hall insulator ($z$-AFQSHI). The results are for the number of bath
   site $n_b=5$ in the ED impurity solver.}
   \label{fig:pd}
   \end{center}
\end{figure}

The phase diagram in Fig. \ref{fig:pd} concludes our findings in this section. The results are for the
spin-orbit coupling $\so=0.2t$. The number of bath sites used in the ED impurity solver is $n_b=5$
although increasing the number of bath sites is not expected to change the phase boundaries.
For small values of the alternating sublattice potential $\delta$ the system is in the
topologically trivial $xy$-AFI which undergoes a transition to the topologically trivial KI
upon increasing the Kondo interaction. For large values of $\delta$ the topologically trivial
$z$-AFI at the weak Kondo coupling and the KI at the strong Kondo coupling
are separated by an intermediate $z$-AFCI. The $z$-AFCI appears around $\delta = 3\jk/4$.
The $z$-AFQSHI is stabilized at moderate values of $\delta$ and the weak Kondo interaction.
The phase diagram in Fig. \ref{fig:pd} modifies and generalizes the previous phase diagrams
\cite{Feng2013,Zhong2013,Yoshida2016} proposed for the half-filled KMK model at zero temperature.

A clarification on the existence of the $z$-AFQSHI at a finite $\delta$ in Fig. \ref{fig:pd} is in order.
The time-reversal symmetry is often considered necessary for the realization of the QSHI protecting the
state against spin-mixing terms such as the Rashba spin-orbit coupling \cite{Kane2005,Wu2006}.
For the AFQSHI the combination of the
time-reversal transformation and a lattice translation is considered to play the role of the protecting
symmetry \cite{Niu2017,Mong2010}.
The QSHI, however, is found to remain immune against the Rashba spin-orbit coupling even in the
presence of a perpendicular Zeeman field perturbation breaking the time-reversal symmetry \cite{Yang2011a,
Li2012,Li2013,Luo2017}.
The $z$-AFQSHI
appearing at a finite $\delta$ in Fig. \ref{fig:pd} is also expected to be robust against the spin-mixing terms
despite the absence of a lattice translation to compensate the effect of the time-reversal transformation
\cite{Guo2011}.
There is usually a sub-symmetry in the system protecting the edge states \cite{Wang2023a} although finding it is not always
trivial \cite{Luo2017}.
An explicit analysis of the effect of the Rashba spin-orbit coupling on the edge states in the $z$-AFQSHI
and the existence of a certain hidden symmetry deserve future attention.

The stabilization of the $z$-AFQSHI and the $z$-AFCI in Fig. \ref{fig:pd} is a consequence
of the combined effect of the spin-orbit coupling, the electronic correlation, and the
alternating sublattice potential. The $z$-AFCI is reported also in some other models of
strongly correlated systems under similar
conditions but for transition-metal compounds \cite{Jiang2018,Ebrahimkhas2022,Guo2023,HafezTorbati2024}.
The Mott insulator phase in the strongly interacting regime
of these models is always magnetically ordered in contrast to the paramagnetic KI phase in Fig. \ref{fig:pd}
which is the characteristic of heavy-fermion systems.

\section{Charge Excitations}
\label{sec:gap}
This section is devoted to a comprehensive study of charge excitations in the KMK model with a three-fold aim.
First, we confirm that for the topologically trivial phases in Fig. \ref{fig:pd} the charge excitations are gapped
in the bulk and at the edges but for the $z$-AFCI there are gapless excitations localized at the edges for only
one spin component. This demonstrates the topological nature of these phases beyond the topological Hamiltonian method.
Second, we examine that the topological phase transition from the trivial $z$-AFI to the $z$-AFCI predicted based
on the topological Hamiltonian is accompanied by closing of the bulk charge gap in the KMK model.
Finally, we address the dependence of the charge gap on the spin-orbit coupling in the $z$-AFCI.
This allows a systematic comparison between the size of the charge gap in the AFCI in heavy-fermion systems and
the size of the charge gap in the AFCI reported for transition-metal compounds \cite{HafezTorbati2024}.

To investigate the charge excitations in the bulk and at the edges we consider a cylindrical geometry of the
size $L\times L$ with $L=40$. We consider the periodic boundary condition in $y$ and the open boundary condition
in $x$ direction with the edges at $x=0$ and $x=39$. We compute the single-particle spectral function
\bearr
A_x(\omega)&:=&\frac{1}{2} \sum_{\alpha=\up,\dn} A_{x;\alpha}(\omega) \nn \\
&:=&\frac{1}{4} \sum_{\alpha=\up,\dn} \left( A_{x,y;\alpha}(\omega)+A_{x,y+1;\alpha}(\omega) \right)
\label{eq:sf}
\eearr
where $A_{x,y;\alpha}(\omega)$ is the local spectral function at the lattice position $(x,y)$ for spin $\alpha$.
The Lorentzian broadening with a broadening factor of $0.05t$ is used in the calculations.
In Eq. \eqref{eq:sf} we have averaged over the spin and the two non-equivalent lattice sites in the $y$ direction.
The spin-resolved spectral function $A_{x;\alpha}(\omega)$ is utilized in the case we aim to distinguish
the contributions of different spins to the spectral function.

\begin{figure}[t]
   \begin{center}
   \includegraphics[width=0.7\textwidth,angle=-90]{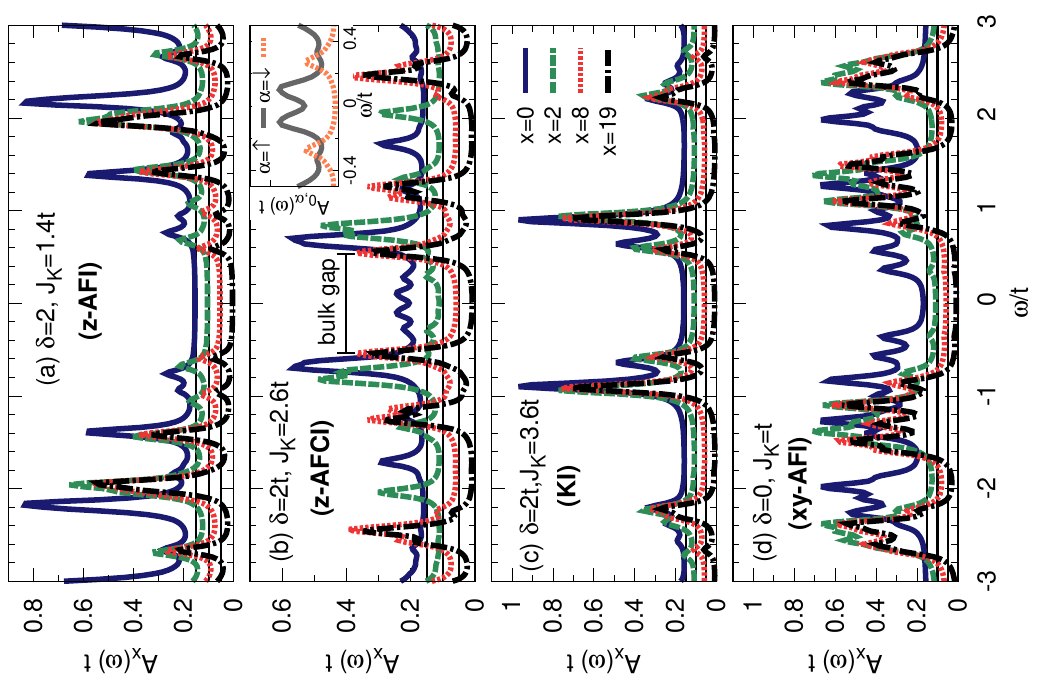}
   \caption{Single-particle spectral function \eqref{eq:sf} plotted vs frequency $\omega$
   around the Fermi energy $\omega=0$ at various values of $x$
   for a cylindrical geometry of size $40\times 40$ with periodic boundary conditions in
   $y$ and open boundary conditions in $x$ direction. The edges are at $x=0$ and $x=39$. The spin-orbit
   coupling $\so=0.2t$ and the number of bath sites in the ED impurity solver $n_b=5$.
   The different panels represent results for different values of the alternating sublattice potential $\delta$
   and the Kondo interaction $\jk$, which are chosen to correspond to the different phases in the phase diagram in Fig. \ref{fig:pd}.
   The horizontal bar in the panel (b) denotes the bulk gap and the inset exhibits the edge spin-resolved spectral function
   $A_{x=0,\alpha}(\omega)$.}
   \label{fig:sf}
   \end{center}
\end{figure}

The spectral function \eqref{eq:sf} is plotted in Fig. \ref{fig:sf} around the Fermi energy $\omega=0$ for $\so=0.2t$
and different values of the alternating sublattice potential $\delta$ and the Kondo interaction $\jk$
corresponding to the different phases $z$-AFI, $z$-AFCI, KI, and $xy$-AFI.
We find the spectral function $A_x(\omega)$ perfectly symmetric with respect to the center of the cylinder
which is why we have included only the results for $x<20$ in Fig. \ref{fig:sf}. The spectral function $A_x(\omega)$
becomes independent of $x$ as we move away from the edges. This can be seen in all the panels in Fig. \ref{fig:sf}
as the results for $x=8$ nicely agree with the results for $x=19$, matching perfectly also with the spectral function
that we obtain by applying periodic boundary conditions in both directions.

As one can see from Figs. \ref{fig:sf}(a), (c), and (d) the charge excitations in the topologically trivial phases
are gapped in the bulk and at the edges.
In contrast, there is a finite spectral weight at the Fermi energy $\omega=0$
in the $z$-AFCI in Fig. \ref{fig:sf}(b) for $x=0$ which quickly disappears as the bulk is
approached. The spin-resolved spectral function $A_{x=0,\alpha}(\omega)$ depicted in the inset of
Fig. \ref{fig:sf}(b) verifies that the gapless edge excitations is due to the spin $\up$ which is the
one entering the topological region in Fig. \ref{fig:delta2}(c).
This demonstrates the existence of
the $z$-AFCI directly for the interacting model \eqref{eq:kmk} going beyond the topological Hamiltonian method.

\begin{figure}[t]
   \begin{center}
   \includegraphics[width=0.31\textwidth,angle=-90]{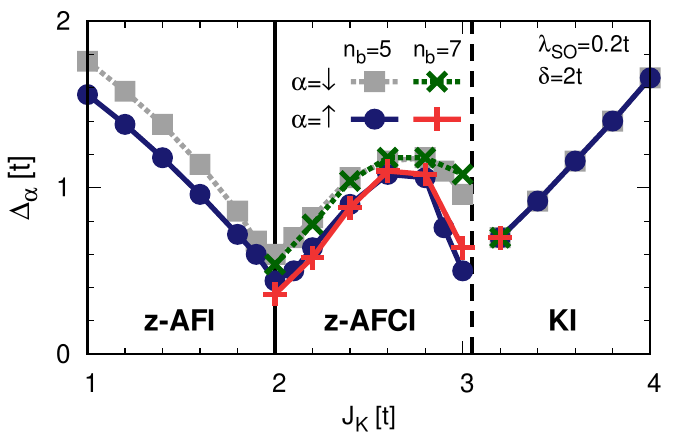}
   \caption{Spin-resolved charge gap vs the Kondo interaction $\jk$
   for the alternating sublattice potential $\delta=2t$ and the spin-orbit coupling $\so=0.2t$.
   The vertical solid line denotes the position of the transition point from the topologically trivial $z$-AF insulator
   ($z$-AFI) to the $z$-AF Chern insulator ($z$-AFCI) predicted by the topological Hamiltonian.
   The vertical dashed line depicts the position of the first-order transition point from the $z$-AFCI to the
   Kondo insulator (KI).
   In the intermediate region the results for the number of bath sites
   $n_b=5$ and $7$ are compared.}
   \label{fig:gap}
   \end{center}
\end{figure}

The topological phase transition from the trivial $z$-AFI to the $z$-AFCI predicted by
the topological Hamiltonian upon increasing the Kondo interaction $\jk$ in Fig. \ref{fig:delta2}(c)
has to be accompanied by the softening of the charge gap in the KMK model \eqref{eq:kmk}. This is necessary
for the Chern number to change across the phase transition.
In Fig. \ref{fig:gap} we have plotted the charge gap $\Delta_\alpha$ for spin $\alpha$ vs the Kondo interaction $\jk$ for
the alternating sublattice potential $\delta=2t$. The spin-orbit coupling is set to $\so=0.2t$.
The charge gap $\Delta_\alpha$ is extracted from the spin-resolved
spectral functions obtained using periodic boundary conditions in both directions for a lattice of size $L \times L$
with $L=40$.
The charge gap is given by the distance between the electron and the hole peaks closest to the Fermi energy $\omega=0$
in the spectral function.
This is illustrated by a horizontal bar in Fig. \ref{fig:sf}(b).
The charge gap is defined as the energy required to add an electron plus the energy required to remove an
electron from the system. It is sometimes referred to also as the band gap.
The results in Fig. \ref{fig:gap} are for the number of bath sites $n_b=5$
although in the intermediate region we have provided data also for $n_b=7$. We have produced data also for $n_b=6$ in the whole
parameter range showing a nice agreement with the results for $n_b=5$ and not included in Fig. \ref{fig:gap} in order to avoid
a busy figure. As in the previous cases the results are for the $z$-AF solution with the local magnetization of the itinerant
electrons $\langle s_i^z \rangle$ positive on the higher-energy sublattice $A$ and negative on the lower-energy sublattice $B$, which
is why the charge gap for spin $\up$ is smaller than the charge gap for spin $\dn$  in Fig. \ref{fig:gap}.
The vertical solid line in Fig. \ref{fig:gap} specifies the position of the transition point from the trivial $z$-AFI
to the $z$-AFCI predicted by the topological Hamiltonian. The vertical dashed line
shows the location of the first-order transition from the $z$-AFCI to the KI.

The charge gap in the $z$-AFI in Fig. \ref{fig:gap} decreases upon increasing the Kondo interaction up to $\jk=2t$
and starts increasing as the system enters the $z$-AFCI phase.
We attribute the finite value of the charge gap at $\jk=2t$ to the finite number
of bath sites in the impurity solver. One can see a tendency towards a smaller charge gap at $\jk=2t$ upon increasing the number
of bath sites. Capturing a truly vanishing gap at the transition point would require an infinite number of bath sites, i.e.,
a continuous representation of the bath.
We conclude that
the minimum of the charge gap occurring at $\jk=2t$ is nicely in agreement with the prediction of the topological
Hamiltonian for a transition from the $z$-AFI to the $z$-AFCI.
As the effective alternating sublattice potential for spin $\up$ in Fig. \ref{fig:delta2}(c)
approaches the lower boundary of the topological region at $\tilde{\delta}_\alpha\approx -1.04t$ we observe also a decrease
in the charge gap in Fig. \ref{fig:gap} until the first-order transition to the KI happens.
In the KI phase the charge gap becomes spin-independent and increases upon increasing the Kondo interaction as is expected.

\begin{figure}[t]
   \begin{center}
   \includegraphics[width=0.42\textwidth,angle=-90]{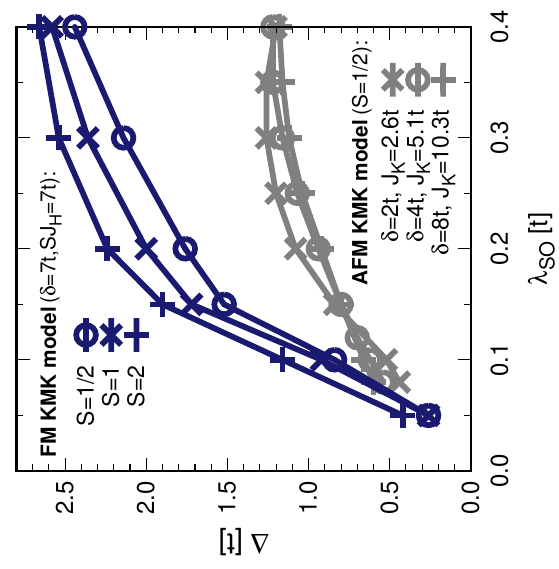}
   \caption{Charge gap for the Kane-Male-Kondo (KMK) model vs the spin-orbit coupling $\so$
   in the $z$-AF Chern insulator ($z$-AFCI) phase. The size of the localized spin is denoted by $S$ and the
   alternating sublattice potential by $\delta$. The figure
   involves the data for both the antiferromagnetic Kondo interaction (AFM KMK model) controlled by $\jk$ and
   the ferromagnetic Kondo interaction (FM KMK model) controlled by $S\jh$. The results are for the
   number of bath sites $n_b=5$ in the ED impurity solver. The data for the FM KMK model are from
   Ref. \onlinecite{HafezTorbati2024}.}
   \label{fig:gap_vs_so}
   \end{center}
\end{figure}

The $z$-AFCI is found in various single-orbital \cite{Jiang2018,Ebrahimkhas2022,Guo2023} and multi-orbital \cite{HafezTorbati2024}
models of transition-metal compounds. The multi-orbital
analysis relies on the KMK model with a {\it ferromagnetic} Kondo interaction (Hund coupling $\jh>0$) between the electron spins
and the localized spins. Different sizes
of the localized spin $S=1/2$, $1$, $2$ corresponding to different numbers of orbitals are addressed \cite{HafezTorbati2024}.
In Fig. \ref{fig:gap_vs_so} we have plotted the charge gap vs the spin-orbit
coupling in the $z$-AFCI phase. The figure involves not only  the data for the KMK model with the AF Kondo interaction
(AFM KMK model) controlled by $\jk$ but also the data for the KMK model with the ferromagnetic Kondo interaction (FM KMK model)
controlled by $S\jh$. The data for the FM KMK model are from Ref. \onlinecite{HafezTorbati2024}.
The results in Fig. \ref{fig:gap_vs_so}
allows a systematic comparison between the size of the charge gap in the $z$-AFCI in heavy-fermion materials
and in transition-metal compounds.

The values for the alternating sublattice potential and the Kondo interaction in Fig. \ref{fig:gap_vs_so}
are chosen such that the system falls
in the AFCI phase. We see only negligible changes in the charge gap in the AFM KMK model despite
the significant
change in the alternating sublattice potential $\delta$. This shows that the results do not depend on the details
of the model and represent the generic behavior of the charge gap in the AFCI in heavy-fermion systems.
No significant change in the charge gap is observed also in the FM KMK model for various sizes of the
localized spin and hence the results for the FM KMK model are also expected to manifest the generic
behavior of the charge gap in transition-metal compounds.
The results in Fig. \ref{fig:gap_vs_so} unveil the fact that the AFCI exists
in transition-metal compounds with a larger charge gap than in heavy-fermion systems.
This suggests that it is not solely the spin-orbit coupling which determines
the size of the charge gap in an AFCI and other mechanisms are also involved.
The exchange and the double exchange interactions are the responsible mechanisms for the magnetic
blue shift of the charge gap observed in multiple theoretical and experimental studies of AF Mott insulators
\cite{Bossini2020,Hafez-Torbati2021,Hafez-Torbati2022}.
A future research addressing the mechanisms which govern the charge gap in an AFCI
can further pave the path to a higher temperature realization of this intriguing topological
state of matter.

\section{conclusion}
\label{sec:con}
Heavy-fermion materials display a variety of phases such as metals, unconventional superconductors,
trivial and topological Kondo insulators, and different kinds of long-range magnetic orderings.
In this paper we explore the possibility for the realization of
the AFCI and the AFQSHI in heavy-fermion systems. Our analysis is based on
the half-filled KMK model that serves as a paradigm for the study of the effect of the spin-orbit
coupling in heavy-fermion systems.
We go beyond the previous analyses of the model by considering
the N\'eel AF solutions with the local magnetization oriented not only perpendicular, $z$-AF solution,
but also parallel, $xy$-AF solution, to the honeycomb plane. Comparing the energies of the two solutions
reveals that in the absence of the alternating sublattice potential it is the $xy$-AF solution which
is always the stable phase. The model undergoes a transition from the $xy$-AFI to the KI upon increasing
the Kondo interaction $\jk$. No non-trivial topological phase occurs. Our results modify the previous
findings for the phase diagram of the model restricted only to the paramagnetic and
the $z$-AF solutions \cite{Feng2013,Zhong2013,Yoshida2016}.
We address the effect of a finite alternating sublattice potential $\delta$ on the system and map out
the phase diagram of the KMK model \eqref{eq:kmk} in the plane of $\delta$ vs $\jk$.
It is shown that the $z$-AF solution
can become the stable phase at a finite alternating sublattice potential giving rise  to the emergence of
the $z$-AFCI and the $z$-AFQSHI.
Two-dimensional materials with strong spin-orbit coupling and buckled structure such as silicene, germanene,
and stanene (known as 2D-Xenes) \cite{Molle2017} grown on an appropriate magnetic insulating substrate
\cite{Tokmachev2022} are potential candidates to realize
the KMK model with a tunable alternating sublattice potential
and the observation of our predicted AF topological insulators.

We perform a comprehensive study of the charge excitations in the system.
For the topologically trivial phases the charge excitations are confirmed to be gapped in the
bulk and at the edges. In contrast, gapless excitations localized at the edges for only one spin
component appear in the spectrum in the $z$-AFCI phase.
The results for the bulk charge gap across the topological phase transitions exhibit a clear tendency towards a
vanishing charge gap at the location of the topological transition points, confirming further the findings
based on the topological invariant calculations.
A systematic comparison between the size of the charge gap in the AFCI in heavy-fermion materials
and in transition-metal compounds indicates a larger charge gap for the AFCI in the latter
system for the same strength of the spin-orbit coupling. This suggests that there are other mechanisms
in addition to the spin-orbit coupling which determine the charge gap in an AFCI and deserve future attention.
Another major outlook of our work is to explore the topological properties at finite temperatures and
to examine up to which temperature scale the quantization of the Hall conductance and the gapless edge states
can survive.

\section*{references}

%

\end{document}